# Near-Perfect Broadband Infrared Metamaterial Absorber Utilizing Nickel

## HENRIK PARSAMYAN

*Department of Microwave Physics, Yerevan State University, 1 Alex Manoogian, Yerevan 0025, Armenia*
*Corresponding author: hparsamyan@ysu.am*

**We propose a thin, compact, broadband, polarization and angle insensitive metamaterial absorber based on a tungsten reflector, silicon spacer and a top pattern composed of a double square-like ring resonator utilizing nickel (Ni). In such a structure, a high absorption (above 80 %) bandwidth ~ 4.8 μm from 3.52 up to 8.32 μm corresponding to the relative bandwidth ~81% can be achieved with deeply subwavelength unit cell dimensions. Here the physical origin of the broadband absorption is associated with low *Q*-factor dipole modes of the top pattern inner and outer sides functioning as rectangular nanoantennas. Owing to the structural symmetry, the absorber shows a good incidence angle tolerance in the relatively wide range for both TE and TM polarizations. The effective parameters of the Ni-based absorber were retrieved using the constitutive effective medium theory and the absorption characteristics of the effective medium and metamaterial were compared.**

## 1. INTRODUCTION

During past years, metamaterial absorbers (MMA) representing large arrays of subwavelength unit cells and aiming to efficiently absorb most of the incident electromagnetic waves have gained much attention due to several practical applications such as communication, sensing, imaging, energy harvesting, detecting, thermal sources, coding, cloaking etc. [1–7] covering all the electromagnetic spectrum from the visible to the terahertz and microwaves. Depending on the absorption characteristics, MMAs are divided into three main types: single-, multi- and broadband absorbers. The basic design of all absorber types includes a bottom metallic reflector, a dielectric spacer and a top metallic layer [7]. Although efficient absorbers utilizing randomly distributed subwavelength configurations as a top layer have been suggested [8–10], the design of MMAs with specially structured top patterns is predominant [4,7].

One of the important aspects of metamaterials is the absorption bandwidth (BW), which has a great influence on its practical applications. To achieve multiband [11,12] or broadband [13–15] absorption, MMAs with top patterns composed of several resonant subwavelength structures are used, where the absorption band is formed by combining the adjoining resonances of all the resonators. In the optical spectrum, the most widely used metals in MMAs are silver (Ag) and gold (Au) due to their remarkable optical properties and the ability to support surface plasmon modes resulting in strong resonant absorption [16,17]. However, in recent studies, graphene [11,18] and highly lossy metals such as tungsten (W), chromium (Cr), nickel (Ni) and titanium (Ti) have also extensively exploited [19–22]. Abiding by the above strategies of selecting materials and configurations, multiband and broadband absorbers were developed from the optical to microwave frequencies [7,10]. Particularly, in the infrared, many broadband absorbers utilizing surface plasmon phenomenon or electric/magnetic dipole resonances have been demonstrated [14,20,23,24]. However, the BWs of these absorbers are usually about the half of the center wavelength of the absorption band [14,15,25–27] and in order to broaden the absorption BW, unit cells with additional material layers and top resonators are used leading to the larger unit cell dimensions and thicker structures [14,28].

Here, we propose a thin, compact, broadband, polarization and angle insensitive MMA absorber covering most of the mid- and long-wave infrared spectra based on tungsten-silicon (W-Si) reflector-spacer with a double square-like ring top metallic pattern composed of eight non-intersecting rectangular nanoantennas. We show that the efficient (over 80%) broadband absorption with a BW ~ 4.8 μm can be achieved in 3.52-8.32 μm spectrum with a Ni-based [29] top pattern caused by highly lossy optical properties of Ni. Due to the structural symmetry, the absorptance is insensitive to the incident field angle in a relatively wide range for both TE and TM polarizations. Comparison with other mid- and long-wave infrared broadband absorbers reveals that such a broad BW is realized with a thinner and very compact unit cells [14,15,25,27]. The theory of the constitutive effective medium [30] is employed to derive effective material parameters of the Ni-based structure and show the correspondence of the absorption properties of the proposed system and the effective medium.

The areas of potential applications of the proposed device include sensing, detecting, light modulating, cloaking and optical buffering [2–5].

## 2. ABSORBER GEOMETRY AND MODELLING

The structure of the considered absorber is composed of a metal reflector substrate, a middle dielectric spacer and a second metallic layer, schematic 3D view of which is shown in Fig. 1(a). Lateral cross-section of the absorber unit cell is shown in Fig. 1(b).

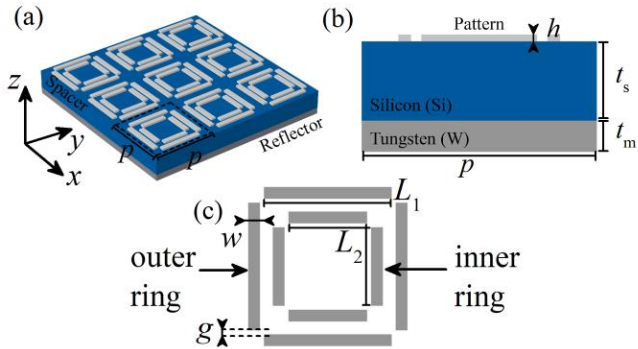

Fig. 1. (a)-(b) 3D and 2D view of the absorber structure: $p$ is the periodicity of the unit cell in both $x$ and $y$ directions, $t_s$ – thickness of the Si substrate, $t_r$ – the thickness of the bottom W reflector and $h$ – the thickness of the top metal structure. (c) The configuration of the double square-like rings. $L_1$ and $L_2$ are lengths of outer and inner strip lines, respectively, $w$ is the width of lines and $g$ – the gap between them.

$t_m$ is the thickness of the bottom reflector, $t_s$ – of the Si spacer, $h$ – of the top patterned structure and $p$ is the periodicity of the absorber unit cell in both $x$ and $y$ directions. Fig. 1(c) depicts the proposed double square-like ring resonator patterned structure for the near-perfect absorption consisting of rectangular nano antennas as sides of square-like rings: $L_1$ and $L_2$ are lengths of outer and inner ring sides, respectively. $g$ is the gap between corners of sides and $w$ – their widths.

Numerical analysis based on the finite element method is conducted to investigate the absorber. Within all simulations, $t_m$ = 100-nm-thick W [31] and $t_s$ = 300-nm-thick Si [32] were used as a bottom reflector and an insulator spacer, respectively. The broadband absorption of the device is studied by using Ni [29] as a metal for the top pattern. The unit cell of the absorber with the periodicity of $p$ = 1 μm was modelled by applying Flaquet periodic boundary conditions on side walls perpendicular to $x$ and $y$ directions. The absorptance can be calculated by a general expression $A(\lambda) = 1 - R(\lambda) - T(\lambda)$, where $R(\lambda) = |S_{11}|^2$ and $T(\lambda) = |S_{21}|^2$ are wavelength-dependent reflection and transmission, expressed through simulated scattering $S$-parameters. In the wavelength range of interest from 1.2 up to 10 μm, the skin depth of the W reflector is much smaller than its $t_m$ thickness, thereby the transmission can be neglected $T(\lambda) \sim 0$ and the relation for the absorptance is simplified to $A(\lambda) = 1 - R(\lambda)$. Throughout all the analysis the absorptance will be considered as high if it is equal to or greater than 0.8 [14,26] (horizontal red dashed lines in Fig. 2). The relative bandwidth (RBW) of the absorption is defined as the ratio of the absorption bandwidth to the center wavelength $\lambda_c$ of the absorption band (×100% in the percentage). The absorber can be easily fabricated by common nanofabrication methods utilizing physical deposition and conventional lithography techniques.

## 3. RESULTS AND DISCUSSION

We analyzed the absorption performance of the proposed W-Si-based metamaterial absorber (MMA) with a top pattern composed of double square-like ring resonators depicted in Fig. 1(c) utilizing Ni.

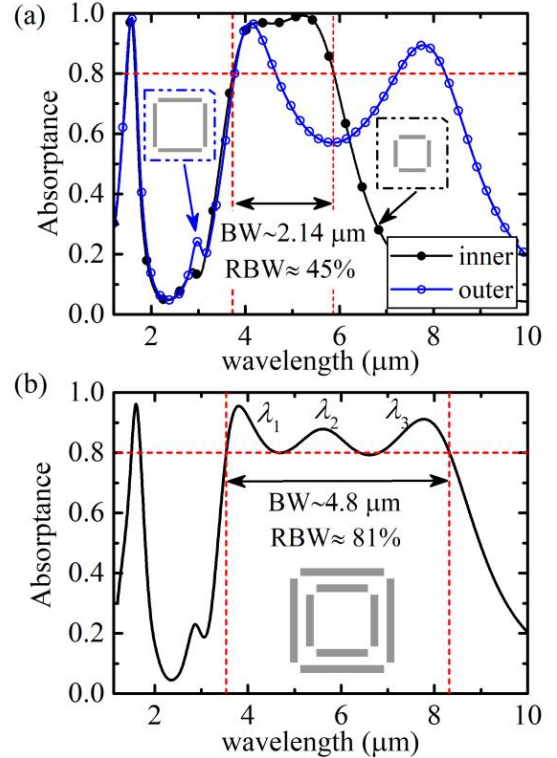

Fig. 2. Simulated absorptance of the metamaterial absorber with (a) inner (black) and outer (blue) square-like ring only and (b) double square-like Ni rings as a top structured pattern. $w$ = 75 nm, $L_1$ = 650 nm, $L_2$ = 400 nm, $g$ = 20 nm and $h$ = 30 nm. RBW represents the absorption bandwidth as a percentage of the center wavelength of the high absorption range.

To describe the contribution of either inner and outer rings in the absorption characteristics of the MMA with double square-like ring resonators and to reveal the effect of the side length of nanoantennas, in Fig. 2(a) we compared the absorption spectra of the single Ni square-like ring MMA with side lengths $L_1$ = 650 nm (rings) and $L_2$ = 400 nm (solid circles), corresponding to the outer and inner rings of the double square-like ring resonator [see Fig. 1(c)]. While in the 3-10 μm spectrum the structure with the inner ring shows high absorption BW ~ 2.14 μm from 3.73 to 5.87 μm (RBW~45%), the structure with the outer ring has two high absorption peaks at 4.11 μm and 7.78 μm with corresponding values of absorptance ~0.97 and 0.89. The absorption spectrum of the double square-like ring absorber composed of inner and outer Ni rings is shown in Fig. 2(b). Here, efficient absorption appears in the range 3.52-8.32 μm (above 80%) with the bandwidth ~4.8 μm, corresponding to the RBW value of 81% (the center wavelength $\lambda_c$~5.92 μm). Three absorption peaks occur at $\lambda_1$ = 3.8 μm, $\lambda_2$ = 5.6 μm and $\lambda_3$ = 7.78 μm with absorptances ~ 0.95, 0.88 and 0.91, respectively.

In Fig. 2 one notes the resonant peak around 1.56 μm with the absorptance ~ 0.96. The reason for such a high absorptance can be understood from Fig. 3(a), where we illustrated the distributions of the normalized electric and magnetic field magnitudes in $XY$ - plane of the MMA illuminated by a normal incident $x$-polarized plane wave. Arrows in the electric field colormaps (first column) represent the electric field polarization and arrows in the magnetic field colormaps (second column) – currents. One sees that a strong electric field is concentrated within the inner and outer ring sides perpendicular to the incident field polarization, where a capacitor with a distance between plates of 50 nm is formed. Our analysis reveals that such a resonant peak of absorption ~0.9 appears even if there is no top metallic pattern and the structure is only composed of a 100 nm-thick W substrate and a 300-nm-thick Si-layer, which can be attributed to the Fabry-Perot modes of the structure, where the near-perfect absorption occurs due to the interference between the incident wave, the waves reflected either from the Air-Si boundary and the bottom W-plate [33]. The refractive index of Si at the resonant wavelength is $n_{Si}$ = 3.47.

To reveal the physical mechanisms of the broadband absorption of the MMA with double square-like Ni rings, in Fig. 3(b-c) are illustrated the simulated normalized electric ($|E|/E_0$) and magnetic ($|H|/H_0$) field distributions of three high absorption peaks at $\lambda_1$ = 3.8 μm, $\lambda_2$ = 5.6 μm and $\lambda_3$ = 7.78 μm in the $XY$- plane illuminated by a normal incident $x$-polarized electric field. Here arrows in the electric field colormaps (first column) stand for the electric field polarization and arrows in the magnetic field colormaps (second column) stand for currents. From Fig. 3 (b-c) one sees that the absorption resonances are associated with the dipole modes of the top pattern sides parallel to the incident electric field polarization, functioning as rectangular nanoantennas [3,34,35]. Particularly, the first resonance at 3.8 μm corresponds to the dipolar resonance of the inner ring. However, similar to the $\lambda_0$ resonance, here a part of the electric field is concentrated within the inner and outer ring sides along the $y$-axis [see Fig. 3(b)] causing also current flow by the outer ring sides. The second resonance at 5.6 μm is conditioned by the dipolar modes of both inner and outer ring sides, accompanied by currents along these nanoantennas, as can be seen from Fig. 3(c). Finally, the third resonance at 7.78 μm is caused by the dipole mode of the outer ring only, as shown in Fig. 3(d). It is important to mention that such wide dipole resonances of the sides of rings forming an efficient broad absorption band are conditioned by the highly lossy characteristics of the Ni in the infrared such that the real and imaginary parts of the dielectric permittivity are of the same order of magnitude. For instance, at the highest absorption peaks the values of the dielectric permittivity of Ni are $-184.69+140.32i$ at 3.8 μm, $-395.43+223.72i$ at 5.6 μm and $-749.24+392.13i$ at 7.78 μm [29].

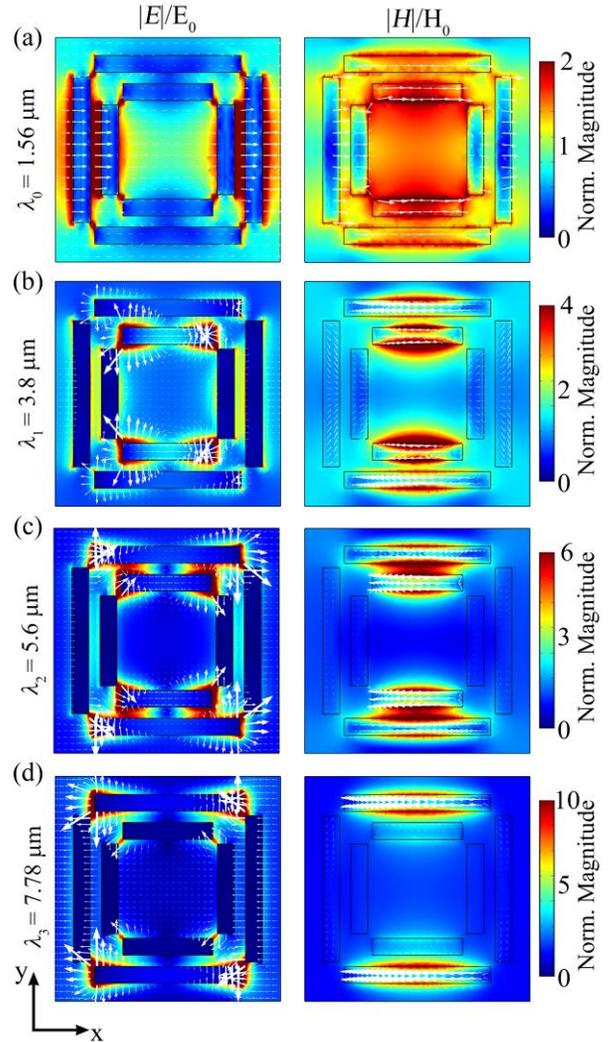

Fig. 3. First column – Electric field magnitude (colormaps) and polarization (arrows) and second column – magnetic field magnitude (colormap) and current distribution (arrows) of the top pattern at absorption peaks (a) $\lambda_0$ = 1.56 μm, (b) $\lambda_1$ = 3.64 μm, (c) $\lambda_2$ = 5.12 μm and (d) $\lambda_3$ = 7.47 μm in $XY$-plane.

From the standpoint of practical applications, it is important to realize an absorber, whose performance is insensitive to the polarization and angle of the incident wave. Fig. 4(a) illustrates the schematic of the studied MMA absorber with oblique incident TM- and TE-polarized plane waves, where the angle between the wavevector $k$ and surface normal $\varphi$ is the incidence angle. All geometrical parameters were chosen as in Fig. 2. One sees that for the TM-polarization, the broad high

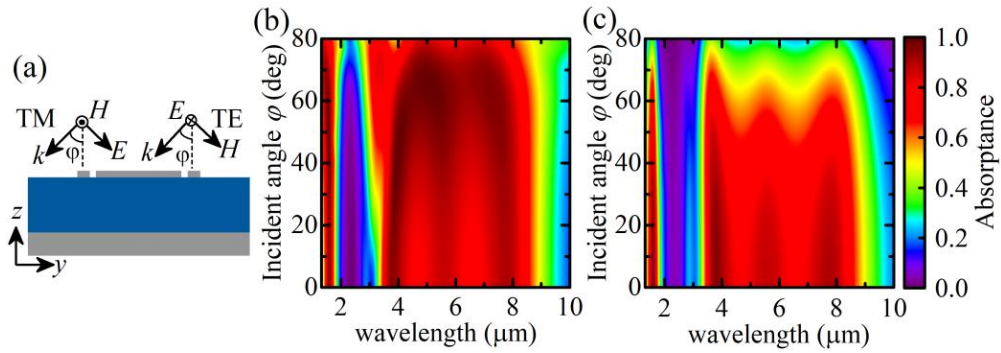

Fig. 4. (a) Schematic of the absorber under an oblique incident TM and TE plane waves. $\varphi$ is the angle between wavevector and surface normal. (b-c) Contour plots of the absorption spectra of the absorber with double square-like Ni rings versus the incident angle $\varphi$ for (b) TM and (c) TE polarized plane waves.

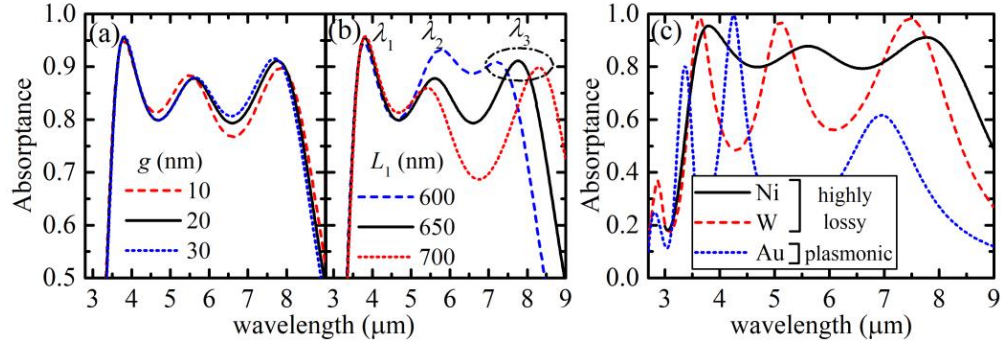

Fig. 5. Dependence of the absorption spectrum on the various values of the (a) gap width $g$ (10, 20, 30 nm) and (b) side length $L_1$ of the outer square-like ring (600, 650, 700 nm). (c) The absorption spectra of the MMA based on the double square-like ring resonators made of Ni, W and Au. All parameters are as in Fig. 2(b).

absorption band about 4.8 μm is preserved for incident angles up to 65°.

Although the absorption bandwidth decreases as a further increase in $\varphi$ starting from ~65°, the high absorptance with the bandwidth of ~2.94 μm in the range from 4.54 to 7.34 μm at $\varphi=80°$ is still achieved (RBW~49%) [see Fig. 4(b)]. Concerning the case of TE-polarization, the absorption bandwidth is maintained unchanged for the angles of the incident field up to ~40°, after which the absorptance decreases. However, three absorption peaks around the value 0.8 are still observed up to 50°. It is worth to note that wide angle tolerance both TE and TM polarizations is due to the applied symmetry in the structure.

The optimization of the absorber was carried out to achieve a high absorptance in the relatively broad absorption band. First, we investigated the absorptance of the structure with only one square-like ring and varied the thickness and width of rectangular nanoantennas, as well as the side length similar to Fig. 2(a) to realize a relatively broad absorption band. After we built another ring encircling the first one and, by fixing the inner ring parameters and the gap size $g$, changed the side length of the outer ring to engineer a broad bandwidth. The influence of the gap size $g$ on the performance of the MMA, when all other parameters are fixed, is shown in Fig. 5(a). One sees that variations in $g$ from 10 nm to 30 nm practically do not affect either the absorption BW and the overall absorptance thereby demonstrating the independence of the absorptance on the possible fabrication errors of the gap size. The dependence of the absorption spectrum on the side length $L_1$ of the outer square-like ring varying from 600 to 700 nm by a 50-nm-step is demonstrated in Fig. 5(b). As was previously noted, the absorption is related to the excitation of dipole modes of sides of rings. Since the inner ring sizes are constant, the first peak $\lambda_1$ corresponding to the dipole mode of the inner ring is unaffected, whereas the red-shift of the $\lambda_3$ is observed by increasing $L_1$, which also yields to the weaker coupling to the $\lambda_2$ mode due to the increase in the distance between the rings since the gap size $g$ is constant.

As the basic material for the top pattern, we chose a highly lossy Ni. However, to compare the performance of the absorber for other metals, in Fig. 5(c) we show the absorptance of the structure in case of the top pattern utilizing another highly lossy metal–W, as well as a common plasmonic metal–gold (Au) [29]. One sees that for the MMA based on either W or Au triple-band absorption spectra occur. Moreover, the absorption band in the case of W is characterized by three resonances at 3.64 μm, 5.12 μm and 7.47 μm with corresponding *near-unity* values of absorptances 0.99, 0.96 and 0.98, respectively, whereas in the case of Au only one near-perfect absorption peak is observed at 4.25 μm, and the resonances at 3.37 μm and 6.95 μm have values 0.8 and 0.62. Relatively narrow resonant peaks of W- and Au-based absorbers are attributed to the lower values of the ratio of the real and imaginary parts of dielectric permittivities of those metals. For instance, at ~ 3.6 μm (near the first resonances) the dielectric permittivities of Ni, W and Au are $-166.84+131.5i$, $-242.83+58.83i$ and $-673.7+92.23i$, respectively.

Table 1. Comparison of the proposed absorber with other mid- and long-wave infrared broadband absorbers

| Ref. | 80% absorption band (μm) | Bandwidth | | Unit cell sizes | | Layers |
|---|---|---|---|---|---|---|
| | | Absolute (μm) | Relative (%) | Absolute (μm$^3$) | Relative to $\lambda_c$ | |
| [15] | 3.2 – 4.4 | 1.2 | 31.6 | 0.35 × 2 × 2 | 0.09 × 0.53 × 0.53 | 3 |
| [25] | 8 – 12 | 4 | 40 | 1.44 × 2 (2D) | 0.14 × 0.25 | 3 |
| [27] | 4 – 6.6 | 2.6 | 49 | 0.68 × 8 × 8 | 0.12 × 1.5 × 1.5 | 3 |
| [14] | 7.7 – 12.2 | 4.5 | 45 | 0.79 × 6.76 × 6.76 | 0.08 × 0.68 × 0.68 | 3 |
| | 5.2 – 13.7 | 8.5 | 90 | 1.56 × 9.2 × 9.2 | 0.16 × 0.92 × 0.92 | 5 |
| Our work | 3.52 – 8.32 | 4.8 | 81 | 0.43 × 1 × 1 | 0.07 × 0.17 × 0.17 | 3 |

Such an absorber with multi-band absorption spectrum has a potential for use as a sensor, detector and selective emitter [7,10]. Moreover, the melting point of W~3422°C will provide the device high thermal stability [21].

In Table 1, we compared the absorption and geometrical characteristics such as absorption band above 80%, absolute and relative BWs, unit cell sizes, their ratio to $\lambda_c$ and the total number of structure layers to those of other broadband mid- and long-wave infrared absorbers. One sees that the features of the proposed Ni-based absorber are comparable with those reported in the literature. Moreover, a high bandwidth ~4.8 μm is achieved with deeply subwavelength unit cell dimensions 0.07$\lambda_c$ × 0.17$\lambda_c$ × 0.17$\lambda_c$, where $\lambda_c$ = 5.92 μm, as well as with only three structural layers.

Fig. 6 (a) shows the complex effective dielectric permittivity ($\varepsilon'+i\varepsilon''$) and the magnetic permeability ($\mu'+i\mu''$) of the structure [30] with top double square-like Ni rings. Solid lines depict the real and dashed lines stand for the imaginary parts of constitutive parameters. The effective parameters of the structure are calculated by $\varepsilon = n/Z$ and $\mu = n/Z$, where $n$ and $Z$ are the refractive index and the impedance, respectively, defined using simulated S-parameters as follows:

$$Z = \pm\sqrt{\frac{(1+S_{11})^2 - S_{21}^2}{(1-S_{11})^2 - S_{21}^2}} \quad (1)$$

$$n = \frac{1}{k_0 d}\left\{\left[[\ln(N)]'' + 2\pi m\right] - i[\ln(N)]'\right\}. \quad (2)$$

Here $k_0$ is the free space wavenumber, $d$ – the total thickness of the structure, $m$ – an integer and $N = X \pm i\sqrt{1-X^2}$, $X = 1/2S_{21}(1-S_{11}^2 + S_{21}^2)$ and signs are chosen so that $Z' \geq 0$ and $n'' \geq 0$.

To verify our analysis and intuitively understand the underlying physics of high absorptance of the *entire structure*, we compared the absorptance of the structure with double square-like Ni ring to the equivalent medium with derived constitutive effective parameters [as shown in Fig. 6(a)] having the height of $d$ = 430 nm. The results are illustrated in Fig. 6(b). Red rings correspond to the MMA and the black line – to the effective medium. The inset shows MMA to effective medium conversion. One sees that the results are in good agreement.

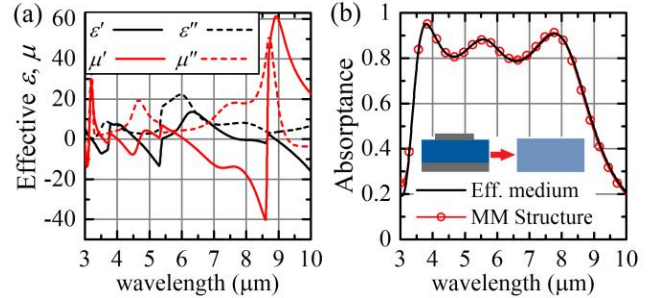

Fig. 6. (a) Real (solid) and imaginary (dashed) parts of the effective dielectric permittivity (black) and magnetic permeability (red) of the MMA with double square-like Ni rings. (b) The absorptance of the metamaterial structure (red-symbols) and the equivalent effective medium (black) with $\varepsilon_{eff}$ and $\mu_{eff}$ shown in (a). All parameters are as in Fig. 2(b).

Moreover, at the first and second absorption peaks both $\varepsilon'$ and $\mu'$ cross zero, while at the third resonance only $\varepsilon'$ crosses zero. Meanwhile, positive resonances of $\varepsilon''$ at the first two absorption peaks mean that these are electric resonances. Note that the results in the range up to 3 μm are not included due to the limitations of the constitutive effective medium theory [30].

## 4. CONCLUSION

In conclusion, we investigated a simple design of an absorber structure based on W reflector, Si spacer and a double square-like ring top pattern composed of non-intersecting rectangular nanoantennas. We showed that efficient broadband absorption over 80% throughout 3.52-8.32 μm range with a bandwidth ~4.8 μm can be achieved by using Ni as a metal for the top pattern, which corresponds to about 81% of the center wavelength of the absorption band. The broadband absorption in the MMA is associated with the low Q-factor dipole resonances conditioned by highly lossy optical properties of Ni. The structural symmetry results in the wide incident angle tolerance up to 65° for TM- and up to 40° for TE-polarized waves. Analyzing the case of tungsten top pattern, one sees that near-perfect triple-band absorption is possible due to lower values of the loss tangent compared to Ni. Comparison to other broadband absorbers in the mid- and long-wave infrared spectra shows that we were able to realize a thinner structure, deeply subwavelength periodicity and at the same time relatively broadband absorption. It is shown that the absorption properties of the proposed system are in good agreement to those of the effective medium, whose properties were retrieved by constitutive effective medium theory.

**Acknowledgment**. We thank Prof. Kh. Nerkararyan and Dr. H. Haroyan for valuable discussions.

**Disclosures.** The authors declare no conflicts of interest.